\documentclass[aps,prl,superscriptaddress,twocolumn,floatfix]{revtex4-2}
\usepackage{units}
\usepackage{amsmath}
\usepackage{amsthm}
\usepackage{amssymb}
\usepackage{graphicx}
\usepackage{color}
\usepackage{xcolor}
\usepackage{bbold}
\usepackage{titlesec} 
\usepackage{enumitem}
\usepackage{pifont}
\usepackage{float}

\definecolor{myurlcolor}{rgb}{0,0,0.7}
\definecolor{myrefcolor}{rgb}{0.1,0,0.9}

\usepackage[
breaklinks,
pdftex,
colorlinks=true, 
linkcolor=myrefcolor,
citecolor=myurlcolor,
urlcolor=myurlcolor
]{hyperref}
\usepackage[capitalize]{cleveref}

\newcommand\bs{{\bar{Y}}}
\def\bbbone{{\mathchoice {\rm 1\mskip-4mu l} {\rm 1\mskip-4mu l}
{\rm 1\mskip-4.5mu l} {\rm 1\mskip-5mu l}}}
\def\CC{{\rm\kern.24em \vrule width.04em height1.46ex depth-.07ex
   \kern-.29em C}}

\DeclareDocumentCommand\mel{ s s m m m }
{ % Matrix element
\IfBooleanTF{#1}
{
\IfBooleanTF{#2}
{\left\langle{#3}\middle\vert{#4}\middle\vert{#5}\right\rangle} % Double starred: total resizing
{\vphantom{#3#4#5}\left\langle\smash{#3}\middle\vert\smash{#4}\middle\vert\smash{#5}\right\rangle} % Starred: no resizing
}
{\vphantom{#4}\left\langle{#3}\middle\vert\smash{#4}\middle\vert{#5}\right\rangle} % Normal: only resize based on bra/ket arguments
}

\newtheorem{lemma}{Lemma}
\usepackage[linesnumbered, ruled,vlined]{algorithm2e}

\graphicspath{{./images/},{./imagesAppendix/}}

\def\app#1#2{%
\mathrel{%
\setbox0=\hbox{$#1\sim$}%
\setbox2=\hbox{%
\rlap{\hbox{$#1\propto$}}%
\lower1.1\ht0\box0%
}%
\raise0.25\ht2\box2%
}%
}

\ifx\proof\undefined
\newenvironment{proof}[1][\protect\proofname]{\par
\normalfont\topsep6\p@\@plus6\p@\relax
\trivlist
\itemindent\parindent
\item[\hskip\labelsep\scshape #1]\ignorespaces
}{%
\endtrivlist\@endpefalse
}
\providecommand{\proofname}{Proof}
\fi

\newtheorem{proposition}{Proposition}
\newcommand{\bra}[1]{\langle #1|}
\newcommand{\ket}[1]{|#1 \rangle}
\makeatother

\newcommand{\ketbra}[2]{|#1 \rangle\!\langle #2 |}

\newcommand{\tr}{\mathrm{tr}}

\providecommand{\factname}{Fact}
\providecommand{\theoremname}{Theorem}
\providecommand{\claimname}{Claim}
\providecommand{\lemmaname}{Lemma}
\providecommand{\definitionname}{Definition}
\providecommand{\corollaryname}{Corollary}
\newtheorem{remark}{Remark}

\definecolor{KB}{rgb}{0.4,0.3,0.9}

\definecolor{THc}{rgb}{0.9,0.3,0.2}

\newcommand{\be}{\begin{equation}}
\newcommand{\ee}{\end{equation}}
\newcommand{\ba}{\begin{equation}\begin{aligned}\hspace{0pt}}
\newcommand{\ea}{\end{aligned}\end{equation}}

\newcommand{\sym}{\text{sym}}

\newcommand{\pur}{\operatorname{Pur}}

\newcommand{\st}[1]{\ketbra{#1}{#1}}
\newcommand{\poly}{\operatorname{poly}}

\newcommand{\sectionMain}[1]{
\let\oldaddcontentsline\addcontentsline% Store \addcontentsline
\renewcommand{\addcontentsline}[3]{}% Make \addcontentsline a no-op
\section{#1}
\let\addcontentsline\oldaddcontentsline
}

\allowdisplaybreaks

\begin{document}

\setcounter{secnumdepth}{3}
%\onecolumngrid
\title{Phase transition in Stabilizer Entropy and efficient purity estimation}

\author{Lorenzo Leone}\email{lorenzo.leone001@umb.edu}%\email{These two authors contributed equally to this paper.}
\affiliation{Physics Department,  University of Massachusetts Boston,  02125, USA}
%\affiliation{Theoretical Division (T-4), Los Alamos National Laboratory, Los Alamos, New Mexico 87545, USA}
%\affiliation{Center for Nonlinear Studies, Los Alamos National Laboratory, Los Alamos, New Mexico 87545, USA}

\author{Salvatore F.E. Oliviero}\email{s.oliviero001@umb.edu}
\affiliation{Physics Department,  University of Massachusetts Boston,  02125, USA}
%\affiliation{Theoretical Division (T-4), Los Alamos National Laboratory, Los Alamos, New Mexico 87545, USA}
%\affiliation{Center for Nonlinear Studies, Los Alamos National Laboratory, Los Alamos, New Mexico 87545, USA}

\author{Gianluca Esposito} \email{gianluca.esposito32@studenti.unina.it}
\affiliation{Dipartimento di Fisica `Ettore Pancini', Universit\`a degli Studi di Napoli Federico II,
Via Cintia 80126,  Napoli, Italy}
\author{Alioscia Hamma}\email{alioscia.hamma@unina.it}

\affiliation{Dipartimento di Fisica `Ettore Pancini', Universit\`a degli Studi di Napoli Federico II,
Via Cintia 80126,  Napoli, Italy}

\affiliation{INFN, Sezione di Napoli, Italy}

\begin{abstract}

Stabilizer Entropy (SE) quantifies the spread of a state in the basis of Pauli operators.  It is a computationally tractable measure of non-stabilizerness and thus a useful resource for quantum computation. SE can be moved around a quantum system, effectively purifying a subsystem from its complex features. We show that there is a phase transition in the residual subsystem SE as a function of the density of non-Clifford resources. This phase transition has important operational consequences: it marks the onset of a subsystem purity estimation protocol that requires $\poly(n)\exp(t)$ many queries to a circuit containing $t$ non-Clifford gates that prepares the state from a stabilizer state. Then, for $t=O(\log_2 n)$, it estimates the purity with polynomial resources and, for highly entangled states, attains an exponential speed-up over the known state-of-the-art algorithms. 
\end{abstract}
\maketitle

%\noindent
\section{Introduction} Quantum information processing promises an advantage over its classical counterpart~\cite{shor1994AlgorithmsQuantumComputation,kitaev1997QuantumComputationsAlgorithms,farhi2016QuantumSupremacyQuantum,boixo2018CharacterizingQuantumSupremacy,harrow2017QuantumComputationalSupremacy,bravyi2018QuantumAdvantageShallow,arute2019QuantumSupremacyUsing}. Since the inception of this field~\cite{gottesman1998HeisenbergRepresentationQuantum}, there has been an extensive theoretical investigation as to what ingredients quantum computation possesses such that it is intrinsically computationally more powerful than classical computation.

 The two resources that set quantum computers apart are entanglement~\cite{bell1964EinsteinPodolskyRosen,bell1966ProblemHiddenVariables,greenberger1990BellTheoremInequalities,page1993AverageEntropySubsystem} and non-stabilizerness~\cite{gottesman1998HeisenbergRepresentationQuantum,bravyi2005UniversalQuantumComputation}. Without either of them, quantum computers cannot perform any advantageous algorithm over classical devices~\cite{shor1994AlgorithmsQuantumComputation,kitaev1997QuantumComputationsAlgorithms,harrow2009QuantumAlgorithmLinear}. In particular, non-stabilizerness measures how many universal gates one can distill from a given quantum state~\cite{campbell2010BoundStatesMagic,campbell2017UnifyingGateSynthesis,campbell2011CatalysisActivationMagic,campbell2017UnifiedFrameworkMagic,bravyi2012MagicstateDistillationLow} and the cost of simulating a quantum state on a classical computer. Indeed, while stabilizer states (the orbit of the Clifford group~\cite{zhu2016CliffordGroupFails}) can be simulated classically in polynomial time~\cite{gottesman1998HeisenbergRepresentationQuantum}, the cost of the simulation scales exponentially in the number of non-stabilizer resources~\cite{bravyi2016ImprovedClassicalSimulation,bravyi2019SimulationQuantumCircuits}, i.e., unitary gates outside the Clifford group~\cite{gottesman1998HeisenbergRepresentationQuantum}.

At the same time, many information processing tasks in quantum computing become inefficient exactly because of the conspiracy of entanglement and nonstabilizerness. Examples of this kind are: state certification~\cite{Leone2023nonstabilizernesshardness}, disentangling~\cite{chamon2014EmergentIrreversibilityEntanglement, shaffer2014IrreversibilityEntanglementSpectrum,yang2017EntanglementComplexityQuantum,zhou2020SingleGateClifford,true2022TransitionsEntanglementComplexity,piemontese2022EntanglementComplexityRokhsarKivelsonsign} or unscrambling~\cite{leone2022RetrievingInformationBlack,leone2022LearningEfficientDecoders,oliviero2022BlackHoleComplexity}. In particular, while purity estimation is a resource-intensive task for universal states, it can be achieved efficiently for stabilizer states. 

It seems then that quantum computation is plagued by a so-called catch-22 dilemma: on the one hand, stabilizer information can be efficiently processed but for the same reason it is useless for a fruitful quantum computation. On the other hand, the combination of entanglement and nonstabilizerness, which makes quantum technology powerful, hinders the efficiency of measurement tasks. Given that, the question posed by this paper is the following: can we leverage the efficiency of information processing offered by the stabilizer formalism for non-stabilizer states?

Recently, a novel measure of non-stabilizerness was introduced as {\em stabilizer entropy} (SE)~\cite{leone2022StabilizerRenyiEntropy}. Stabilizer states have zero stabilizer entropy, whereas non-stabilizer states (those that are computationally useful) exhibit a non-vanishing stabilizer entropy. %This is the entropy of the probability distribution associated with the outcomes from measuring Pauli operators. 
%For pure states, it is also a good measure of $T-$gates distillation. For generally mixed cases, it can be used to obtain strict bounds to the complexity of direct quantum verification protocols~\cite{leone2022MagicHindersQuantum}. 
 Unlike other measures~\cite{campbell2010BoundStatesMagic,howard2017ApplicationResourceTheory,liu2022ManyBodyQuantumMagic}, it is computable (though expensive) and experimentally measurable~\cite{oliviero2022MeasuringMagicQuantum,haug2023ScalableMeasuresMagic,odavic2022ComplexityFrustrationNew}. SE is also involved in the onset of universal, complex patterns of entanglement~\cite{piemontese2022EntanglementComplexityRokhsarKivelsonsign,true2022TransitionsEntanglementComplexity}, quantum chaos~\cite{leone2021IsospectralTwirlingQuantum,oliviero2021RandomMatrixTheory,leone2021QuantumChaosQuantum,oliviero2021TransitionsEntanglementComplexity}, complexity in the wave function of quantum many-body systems~\cite{oliviero2022MagicstateResourceTheory,haug2023QuantifyingNonstabilizernessMatrix}, and the decoding algorithms from the Hawking radiation from old black holes~\cite{leone2022RetrievingInformationBlack,oliviero2022BlackHoleComplexity,leone2022LearningEfficientDecoders}. In the context of operator spreading, it is akin to the string entropy~\cite{chamon2022QuantumStatisticalMechanics}.

As we said, when states possess SE, measurement tasks tend to become inefficient. However, the intriguing aspect of entropy is that it can be transferred from one subsystem to another without altering the total entropy, and thus the total computational power of the system. %Clifford operations preserve the total SE.
This parallels the behavior of a Carnot refrigerator that effectively reduces entropy in a system by transferring it to  the environment, all while keeping the entropy of the universe unchanged. 

In this paper, drawing inspiration from this thermodynamic analogy, we show a general scheme of how to push SE out of a subsystem with Clifford operations, effectively \textit{cooling} the subsystem down from its complex features while preserving the total SE, and explore the consequences and implications of this approach. The two main results of this paper are the following:

(i) There is a phase transition in SE driven by the competition between the creation and spreading of non-Clifford resources versus their localization and erasure.

  (ii)  The localized phase (when a subsystem is successfully cleansed of its nonstabilizerness quantified by SE)  allows for a purity estimation algorithm that, for some cases of interest, obtains an exponential speed-up over all the state-of-the-art known algorithms~\cite{ekert2002DirectEstimationsLinear,islam2015MeasuringEntanglementEntropy,kaufman2016QuantumThermalizationEntanglement,linke2018MeasuringEnyiEntropy,vanenk2012MeasuringMathrmTrEnsuremath,elben2018RenyiEntropiesRandom,brydges2019ProbingRenyiEntanglement}. In practice, this result is obtained by constructing a stabilizer state whose subsystem purity is shown to bound the purity of the desired state.

%By gaining sufficient knowledge about the state and constructing a unitary transformation that preserves the total entropy, the local stabilizer entropy can be successfully transferred across subsystems. However, this process is not always possible and there is a sharp phase transition that separates states whose subsystem stabilizer entropy can be cleansed from those that cannot.
    %\item Whenever a subsystem is successfully cleansed of its complex features, quantified by the stabilizer entropy, the state of the subsystem becomes a stabilizer state, leading to efficient execution of all associated tasks in terms of resource utilization. As an illustrative example, we demonstrate a subsystem purity estimation algorithm that outperforms the state-of-the-art approach with an exponential advantage.

{\em Setup.---} Consider a system of $n$ qubits with Hilbert space $\mathcal H\simeq \CC^{\otimes 2n}\simeq \mathcal H_E\otimes\mathcal H_F$ with dimensions $d_X=2^{n_X}$  with $(X=E,F)$ and $n=n_E+n_F$. Let $\ket{\mathbf{0}}\equiv \ket{0}^{\otimes n}$. To every pure density operator $\psi$ on $\mathcal H$ one can associate a probability distribution $P_\psi$ through its decomposition $\psi =d^{-1}\sum_{P\in\mathbb p} \tr (P\psi)P$  in Pauli operators $P\in \mathbb P$     by
$P_\psi=d^{-1}\tr^2(P\psi)$.  Regardless of the purity of $\psi$, its stabilizer purity is  defined as $\operatorname{SP}(\psi) := \sum_P P_\psi^2$. On the other hand, the purity of the state $\psi$ is given by $\pur(\psi) =\tr(\psi^2)$. Defining the ratio $w(\psi):= d\,\operatorname{SP}(\psi)/\pur(\psi) $, the $2-$R\'enyi SE is given by $M(\psi) = -\log_2 w$~\cite{leone2022StabilizerRenyiEntropy}, while the linear SE is defined as $M_{\operatorname{lin}} = 1-w$. Throughout the paper, we define \textit{stabilizer states}\footnote{Note that the null set of $M(\cdot)$ does not contain convex combinations of stabilizer states.} as those for which $M(\psi)=0$.

Through standard techniques one can write these purities as $\operatorname{SP}(\psi)= d\tr(Q \psi^{\otimes 4})$ and  $\pur(\psi)=\tr(T \psi^{\otimes2})$ where $T$ is the swap operator and $Q=d^{-2}\sum_P P^{\otimes 4}$ is the projector onto the stabilizer code~\cite{zhu2016CliffordGroupFails}. Given the bipartition $E|F$ defined above, one can define the SE associated with the subsystem $X$ as $M_{X}(\psi)\equiv M(\psi_X)$ where the partial states are defined as $\psi_E\equiv\tr_F\psi$ and $\psi_F\equiv\tr_E\psi$. In terms of the $Q,T$ operators the $w$ of the partial state reads
$w(\psi_X)= d_X\tr(Q_{X} \psi^{\otimes 4})/\tr(T_X \psi^{\otimes2})$ where now $Q_X, T_X$ are the $Q$ and swap operator on the subsystem $X=E,F$. Note also that if $\psi$ is a stabilizer state then $M_X(\psi)=0$ for every subsystem $X$~\cite{leone2022StabilizerRenyiEntropy}. However, whether the partial trace is in general a SE-non-increasing map is an open question.

In the following, we are interested in the SE of states $\psi_t$ parametrized by a number $t$ of non-Clifford resources. To this end, we consider  outputs $\psi_{t}\equiv C_{t}\st{\mathbf{0}}C_{t}^{\dag}$ of $t-$doped Clifford circuits $C_t$, that is, Clifford circuits in which $t$ non-Clifford gates have been injected, say $T-$gates~\cite{zhou2020SingleGateClifford,leone2021QuantumChaosQuantum}.  Now, since the Clifford group is very efficient in entangling, the states $\psi_{t}$ are typically highly entangled~\cite{chamon2014EmergentIrreversibilityEntanglement,yang2017EntanglementComplexityQuantum,zhu2016CliffordGroupFails,zhu2017MultiqubitCliffordGroups}. As a result if $n_F\ll n_E$, the partial state $\psi_{tF}$ is very close to the maximally mixed state, which is a stabilizer state with SE equal zero. Absent from this picture is the characterization of stabilizer entropy behavior within partitions when $n_F\ll n_E$. To gain insights into this behavior, it is often necessary to conduct averaging over the Clifford orbit. In pursuit of this objective, we introduce the following lemma, which will prove beneficial throughout this paper.
\begin{lemma}\label{lemma1}
Let $\psi$ be a pure quantum state, $\psi^C\equiv C\psi C^{\dag}$ its Clifford orbit and $\psi^C_E\equiv\tr_F (C\psi C^\dag)$. Then, for $n_f=n\mathfrak{f}$ and $0<\mathfrak{f}<1/2$, 
\be
    \mathbb{E}_C\frac{\operatorname{SP}(\psi_E^C)}{\pur(\psi_E^C)}=\frac{\mathbb{E}_C\operatorname{SP}(\psi_E^C)}{\mathbb{E}_C\pur(\psi_E^C)}\left(1+O\left(2^{-n\frac{1-2\mathfrak{f}}{2}}\right)\right)
\ee
\begin{proof}
    Expanding $\pur(\psi_E^C)$ around the average $\mathbb{E}_C\pur(\psi_E^C)$ one has 
    \be
    \pur(\psi_E^C)=\mathbb{E}_C\pur(\psi_E^C)+\Delta_C\pur(\psi_E^C)\,,
    \ee
    where $\Delta_C\pur(\psi_E^C)\equiv\sqrt{\mathbb{E}_C\pur^2(\psi_E^C)-[\mathbb{E}_C\pur(\psi_E^C)]^2}$, and thus
\be
\mathbb{E}_C\frac{\operatorname{SP}(\psi_E^C)}{\pur(\psi_E^C)}=\frac{\mathbb{E}_C\operatorname{SP}(\psi_E^C)}{\mathbb{E}_C\pur(\psi_E^C)}\left(1-\frac{\Delta_{C}\pur(\psi_E^C)}{\mathbb{E}_C\pur(\psi_E^C)}\right)
\ee
i.e., one has that the average of a ratio is equal to the ratio of the averages up to a relative error that is the relative error of the average purity, i.e., the ratio of the purity fluctuations to the average purity. We can use the fundamental result from~\cite{leone2021QuantumChaosQuantum} that states that the relative error over the Clifford orbit is small as long as the bipartition is not the exact half bipartition, see Eq. $(52)$ of \cite{leone2021QuantumChaosQuantum}. For the purpose of this paper consider $n_F=\mathfrak{f}n$ and $\mathfrak{f}<1/2$, one has
\be
\frac{\Delta_{C}\pur(\psi_E^C)}{\mathbb{E}_C\pur(\psi_E^C)}=O\left(2^{-n\frac{1-2\mathfrak{f}}{2}}\right)
\ee
%Since, in this paper, we work in the limit of large $n$ the above approximation holds.
\end{proof}
\end{lemma}
Thanks to this lemma one can then write, up to an exponentially small relative  error
\be\label{ratioavgE}
\mathbb{E}_C[w(\psi_{E}^{C})]=\frac{d_{E}\mathbb{E}_C\,\operatorname{SP}(\psi_{E})}{\mathbb{E}_C\,\pur(\psi_{E}^{C})}
\ee
and in the same way, one can write 
\be\label{ratioavgF}
\mathbb{E}_C[w(\psi_{F}^{C})]=\frac{d_{F}\mathbb{E}_C\,\operatorname{SP}(\psi_{F})}{\mathbb{E}_C\,\pur(\psi_{F}^{C})}
\ee
From the above lemma, one can indeed show that the average $\mathbb{E}_C$ (over the Clifford group) SE in $F$ is very small while it is all contained in the subsystem $E$:
\begin{proposition}\label{prop1} 
Consider $n_{F}/n<1$. Let $\psi$ a pure state and denote $\psi^C\equiv C\psi C^{\dag}$, for $C$ a Clifford circuit. The average over the Clifford orbit of the partial SEs of subsystems $E,F$ is given by 
  \ba
      \mathbb{E}_C[M_{\operatorname{lin}}(\psi^C_E)]&=M_{\operatorname{lin}}(\psi)+O\left(\frac{d_F}{d_{E}}\right)\\
      \mathbb{E}_C[M_{\operatorname{lin}}(\psi^C_F)]&=O\left(\frac{d_F}{d_{E}}\right)
    \ea
for large $n$. See App~\ref{App:proofprop1} for the proof.
\end{proposition}
In the paper, we frequently make use of the big-O notation, see Appendix~\ref{App: bigO} for a brief review.

The above formulas show how the SE is all in the larger system. As a corollary (by plugging the above formulas), on average over the Clifford orbit of $\psi$, the partial trace preserves SE in the sense that $M(\psi)\simeq M_{E}(\psi)+ M_{F}(\psi)$.

\section{Cleansing algorithm} As we have seen, typically all the SE is contained in the greater of the two subsystems, namely the subsystem $E$. As such, quantum information contained in the state $\psi_E$ cannot be efficiently processed~\cite{bravyi2016ImprovedClassicalSimulation}: it contains almost all of the quantum complexity induced by the circuit $C_t$. One asks whether it is possible to {\em cleanse} $E$ from this complexity. If one could do that, one would be able to manipulate $E$ efficiently by means of the stabilizer formalism~\cite{aaronson2004ImprovedSimulationStabilizer}.  Since the SE is an entropy, one wonders if one could toss it in the other subsystem $F$ by means of a suitable protocol.
Here, we set the problem up in the following way: can we find a quantum map $\mathcal{E}(\cdot):= \tr_Y W(\cdot ) W^\dag$ with $W$ a Clifford unitary and $Y$ a subset of $n_Y$ qubits $Y\subseteq F$,  
such that $M_{E}[\mathcal{E} (\psi_t)]=0$?

To this end, we utilize a fundamental result obtained in~\cite{leone2022LearningEfficientDecoders}: given a $t-$doped circuit $C_t$ the so-called \emph{Clifford Completion algorithm} can learn [by  $\operatorname{poly}(n)\exp (t)$ query accesses to a unknown $C_t$]  a  Clifford operator ${D}$ called {\em diagonalizer} such that 
   $C_t={D}^{\dag} c_t {D}V$, where $c_t$ is a $t-$doped Clifford circuit acting on a subsystem with only $t$ qubits (for $t\le n$) and $V$ is another suitable Clifford unitary operator. This result ensures that   $t=O (\log_2 n)$ the operator $D$ will (with negligibly small failure probability) be found in polynomial time.

Now, consider the permutation operator $T_\pi$ on $n$ qubits, namely $T_{\pi}\ket{x}=\ket{\pi(x)}$ with $\pi \in S_{n}$ the symmetric group of $n$ objects. Note that the  $T_\pi$'s also belong to the Clifford group\footnote{Any permutation of the qubits belongs to the Clifford group because the permutation group $S_n$ is generated by swaps operator and the swap operator $S_{ij}$ between the qubits $i$-th and $j$-th is made out of $3$ CNOTs.}. Then, one can choose a suitable permutation $\pi_Y$ such that the dressed operator 
$T_{\pi_Y}c_tT_{\pi_Y}^{\dag}\equiv c_t^Y$ acts non trivially only on any desired subsystem $Y\subset E\cup F$, where $n_Y= t$. In particular, one can choose $Y\subset F$ if $n_Y\le n_F$. 
As a result, one has $T_{\pi_Y}D\ket{\psi_t}=T_{\pi_Y}c_tT_{\pi_Y}^{\dag}\ket{\Phi}=c_t^Y\ket{\Phi}$ with $\ket{\Phi}:= T_{\pi_Y}DV\ket{\mathbf{0}}$ being a stabilizer state (with density operator $\Phi\equiv \st{\Phi}$). 
Then, by picking $W\equiv T_{\pi_Y}D$, we define the Clifford map $\mathcal{E} (\cdot):= \tr_Y W (\cdot) W^{\dag}$ 
that first localizes the non-Clifford resources in $Y$ and then erases them by tracing $Y$ out. We can now prove the following.

\begin{proposition}\label{prop2}
For $ n_Y\le n_F$, the map $\mathcal{E}$ moves the $t$ non-Clifford gates in the subsystem $Y\subset F$, and by tracing out $Y$ makes  the SE on $E$ zero, i.e., $M_E[\mathcal{E}(\psi_t)]=0$.
\begin{proof}
Start with
\ba
M_E[\mathcal{E}(\psi_t)]=-\log_2 d_E\frac{\tr[Q_E\mathcal{E}(\psi_t)^{\otimes 4}]}{\tr[T_{E}\mathcal{E}(\psi_t)^{\otimes 2}]}
\ea
and recall that $\mathcal{E}(\psi_t)=\tr_Y (c^Y_t\Phi c^{Y\dag}_t) $. Now, since $[c_t^{Y\otimes 4}, Q_{E}]=[c_t^{Y\otimes 2}, T_E]=0$, one obtains $\tr[Q_E\mathcal{E}(\psi_t)^{\otimes 4}]= \tr_E (Q_E\Phi_E^{\otimes 4})$ and $\tr[T_E\mathcal{E}(\psi_t)^{\otimes 2}]= \tr_E (T_E\Phi_E^{\otimes 2})$ from which one gets
 $M_E[\mathcal E( \psi_t)]=M_E(\Phi)=0$, where the last equality follows from  $\Phi$ being a stabilizer state~\cite{leone2022StabilizerRenyiEntropy}. 
\end{proof}
\end{proposition}

\section{Phase transition in SE} We now show that doped Clifford circuits feature a phase transition in SE due to the competition between a term that creates and spreads SE and a term that localizes and then erases it. The first term is the quantum circuit $C_t$. As we saw in the previous section, this unitary operator can be written as $C_t = D^{\dag} c_t D V$. The unitary $c_t$ does insert a number $t$ of non-Clifford gates in a $t$-qubits subsystem, and then the term $D$ spreads them around the entire system. In the large $n$ limit, we can define $\mathfrak{t}:=t/n$ the density of non-Clifford gates and $\mathcal C_{\mathfrak t} $ its adjoint action. The spreading strength of this channel is given by the depth of $C_t$, while its SE strength is given by the density $\mathfrak{t}$.  The map $\mathcal{E}$, on the other hand,  first localizes the SE in the subsystem $Y\subset F$ and then erases it by entangling and tracing out. Thanks to Proposition~\ref{prop2}, its localizing strength is given by the density of qubits belonging to $F$, namely $\mathfrak{f}:=n_F/n$. Altogether, we consider the map composing the two terms, namely $\mathcal E \circ \mathcal C_{\mathfrak t} $ and study the behavior of the SE induced by such a map. In the limit of $\mathfrak{t}/\mathfrak{f}\ll 1$, we expect the map to leave the subsystem $E$ clear of SE. However, for $\mathfrak{t}/\mathfrak{f}\gg 1$, all the SE should be intact in $E$.
\begin{figure}[H]
    \centering
\includegraphics[width=\linewidth]{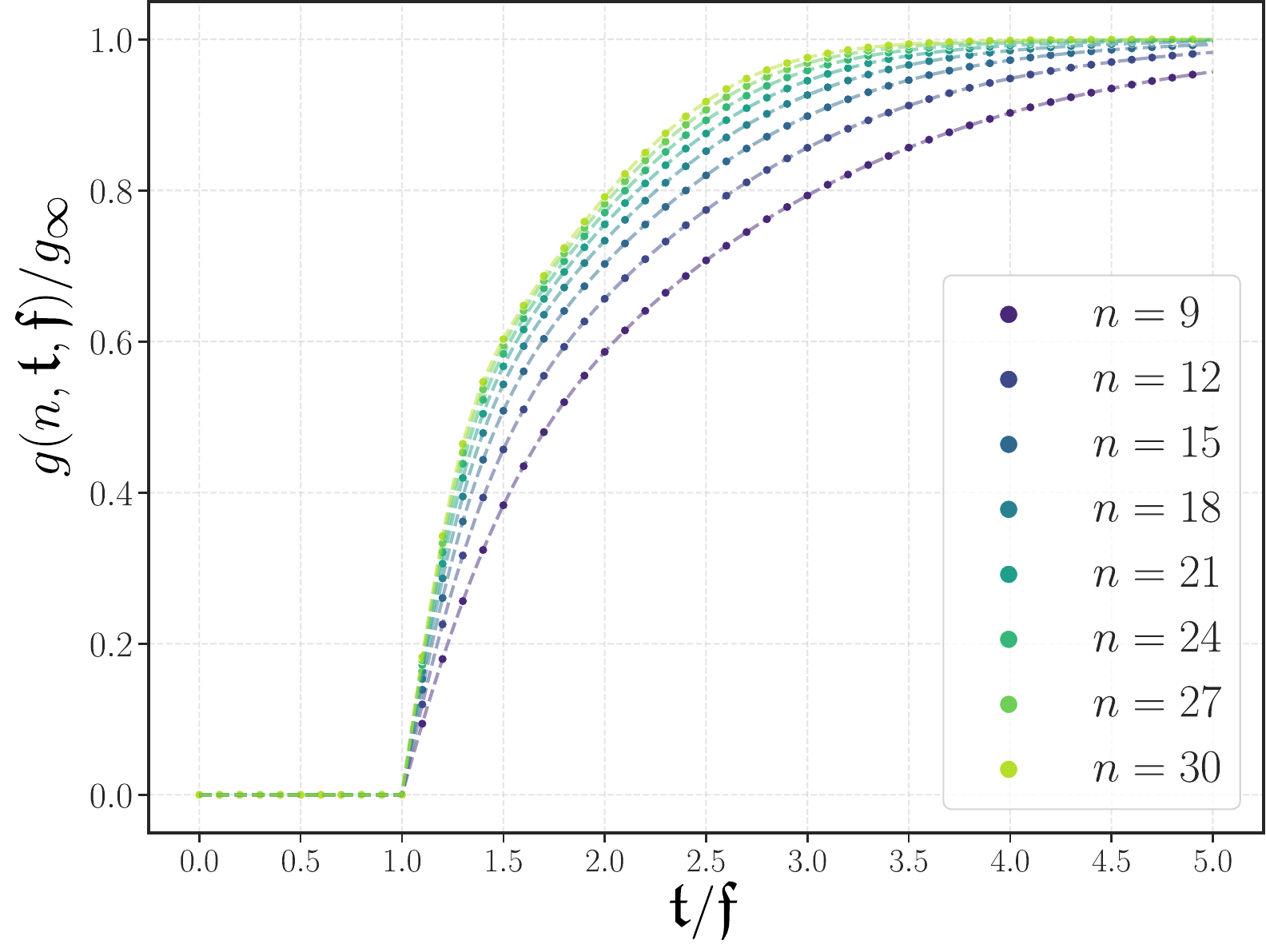}
    \caption{Plot of the ratio between  $g(n,\mathfrak{t},\mathfrak{f})$ and  $g_\infty$, which is a lower bound for the localized SE power of the map $\mathcal{E}\circ \mathcal{C}_{\mathfrak{t}}$, as a function of the ratio of the density $\mathfrak{t}$ of non-Clifford gates to the density $\mathfrak{f}$ of qubits belonging to $F$. We set $\mathfrak{f}=\frac13$. The critical point for the phase transition is $\mathfrak{t}/\mathfrak{f}=1$, with critical index one.}
    \label{pmagt}
\end{figure}

We compute the localizing SE power of the map $\mathcal E \circ \mathcal C_{\mathfrak t} $  by averaging the value of $M_E[\mathcal E \circ \mathcal C_{\mathfrak t} (\omega)]$ over all the maps $\mathcal{E} \circ \mathcal C_{\mathfrak t}$ and all the (pure) stabilizer input states $\omega$. We denote such an average $\mathbb{E}$ (see Appendix~\ref{App:average}). For $\mathfrak{t}/\mathfrak{f}\le 1$ we obtain, in virtue of Proposition 2, $\mathbb{E}\, \left[M_E(\mathcal{E} \circ \mathcal C_{\mathfrak t} [\omega]) \right]=0$. This is the localized phase, where the localizing power of the map $\mathcal E$ prevails. On the other hand, for  $\mathfrak{t}/\mathfrak{f}\ge 1$, direct computation of the average yields, in the delocalized phase,
\ba\label{eq:m2g}
\mathbb{E}  [M_E(\mathcal{E} \circ \mathcal C_{\mathfrak t} [\omega])] \ge g(n,\mathfrak{t},\mathfrak{f})
\ea
See Appendix~\ref{App:average} for details
and the explicit expression for $g(n,\mathfrak{t},\mathfrak{f})$. For large $n$, one has
\be
g_{\infty}\equiv \lim_{t\rightarrow\infty}g(n,\mathfrak{t},\mathfrak{f})\simeq n(1-2\mathfrak{f})
\label{asymptotic}
\ee
In Fig.~\ref{pmagt}, we  plot  $ g(n,\mathfrak{t},\mathfrak{f})/g_{\infty} $ (for  $\mathfrak{f}=1/3$). In the neighborhood of the critical value $\mathfrak{t}/\mathfrak{f}=1$ this ratio behaves as 
$g(n,\mathfrak{t},\mathfrak{f})/g_{\infty}\simeq \frac{\mathfrak{f}}{1-\mathfrak{f}}\left(\mathfrak{t}/\mathfrak{f}-1\right)$, which shows a critical index one, see Fig.~\ref{pmagt}.
%\begin{equation}
%   f(n,t,n_E)= 2^{2(n_F-t)}+O(2^{n-2n_E})
%\end{equation}

\section{Efficient purity estimation} The phase transition described above has relevant operational applications in terms of quantum information processing. We see that in the localized phase $\mathfrak{t}/ \mathfrak{f}\le 1$, the SE can be cleansed from the subsystem $E$, making the subsystem $E$ manipulable by means of the stabilizer formalism. 
We now show that in this phase it is possible to probe the bipartite entanglement in a way that, for cases of interest, gains an exponential speed-up over the state-of-the-art algorithms in the literature~\cite{ekert2002DirectEstimationsLinear,brydges2019ProbingRenyiEntanglement}. 

The best-known way to evaluate the purity of a quantum state within an error $\epsilon$ is the swap test, which requires a number of resources scaling as $O(\epsilon^{-2})$. 
However, typical states possess  a subsystem purity $\Theta(e^{-\tilde{\beta} n_F})$, the so-called {\em volume law} scenario. In this case, to evaluate the purity, one needs to resolve an exponentially small error and therefore exponential (in $n_F$ ) resources, which, since $n_F=\mathfrak{f}n$, is $\Theta(e^{2{\beta} n})$, with $\beta\equiv \tilde{\beta}\mathfrak f$. 
%For the same reason, an experimental witness for volume law of entanglement is exponentially expensive. 

 If the purity one wants to estimate scales polynomially, that is, $\pur(\psi_{E})= \Omega (\mbox{poly}^{-1} (n))$, one will need a polynomial number $N_{\text{shot}}$ of measurements in order to resolve the quantity. Notice that one would not know beforehand what is the number $N_{\text{shot}}$ necessary. In practice, one sets a number of experiments $N_{\text{shot}}$ to obtain  upper bound thresholds $\pur(\psi_{E})\le O(N_{\text{shot}}^{-1/2})$. If $\mbox{poly} (n)$ is a large polynomial, one would in practice be forced to halt the procedure without knowing how tight is the bound. In the worst-case scenario (which is also typical), the purity to evaluate is exponentially small and the estimation will always be exponentially costly.

%Randomized measurement schemes have been shown to have more favorable scalings, although still exponential~\cite{brydges2019ProbingRenyiEntanglement,elben2018RenyiEntropiesRandom}.

We want to show that the cleansing algorithm can give an exponential advantage over the known protocols. The intuition is that if one can cleanse the SE from $\psi_E$, one would obtain a stabilizer state, whose purity can be evaluated with polynomial resources~\cite{fattal2004EntanglementStabilizerFormalism}. 

Let us start with some technical preliminaries.  Consider a state initialized in $\ket{\mathbf{0}}$ and be $\ket{\psi_t} = C_t \ket{\mathbf{0}}$ the output of a $t-$doped Clifford circuit. Its marginal state to $E$ will be denoted by $\psi_E=\tr_F (\psi_t)$. This is the state whose purity we want to evaluate.
 %Notice that $\psi_t$ is typically very entangled. 
 In the localized phase of the cleansing algorithm, we know that the output state $\mathcal{E}(\psi_t)$ is a stabilizer state of $E\cup F\setminus Y$ for $Y\subset F$. Its purity would be easy to evaluate, but it is not directly related to the purity of the original state $\psi_E$ because of the action of $W$.  We now show that we can manipulate the \textit{cleansed state} $\mathcal{E}(\psi_t)$ and construct a stabilizer state $\rho$ whose purity bounds the purity of the desired state $\psi_E$.

 %$\mathfrak{t}/ \mathfrak{f}\le 1$, we know we can write $C_t=W^{\dag}c_t^Y{W}V$ where $c^Y_t$ is a $t-$doped Clifford circuit acting on a subsystem $Y\subset F$ of at most $t$ qubits. By construction, $c_t^Y$ is not entangling in the bipartition $\mathcal H_E\otimes\mathcal H_F$.  

Starting from the cleansed state $\mathcal{E}(\psi_t)$, we first append the maximally mixed state $d_{Y}^{-1}I_Y$ on $Y$ and then we act with the diagonalizer $W^{\dag}$ back, obtaining the state $\rho :=W^{\dag} (\mathcal{E}(\psi_t)\otimes d_{Y}^{-1}I_{Y})W$. Let $\rho_X$ with $X=E,F$ be its marginals and notice that $\rho,\rho_E,\rho_F$ are stabilizer states. What we have effectively done is to re-entangle the state $\mathcal{E}(\psi_t)$ so that it gives a bound to the purity of $\psi_E$.

After these preliminaries, we are ready to establish our protocol. Set $t=O(\log_2 n)$. Utilizing the cleansing algorithm, we first prepare the stabilizer state $\rho$ by learning the diagonalizer $W$, which requires $O(\poly(n))$ resources. Then, since $\rho$ is a stabilizer state, by means of $O(n^3)$ shot measurements, we evaluate $\pur(\rho_X)$ with no error~\cite{fattal2004EntanglementStabilizerFormalism}. Indeed it is sufficient to first learn the stabilizer group $S$ associated to $\rho$ using the algorithm in~\cite{montanaro2017LearningStabilizerStates} and then use the methods developed in~\cite{fattal2004EntanglementStabilizerFormalism} to compute entanglement from the knowledge of $S$ in a computationally efficient fashion. We have two scenarios $(i)$ $\pur(\rho_X)= \Omega (\poly^{-1}(n))$ or $(ii)$ $\pur(\rho_X)=\Theta(2^{-\alpha n})$ and the two following propositions.

%\begin{prop}

\begin{proposition} \label{Prop: exactlowerbound}
The purity of the state $\psi_E$ is lower bounded by
\ba
\pur(\rho_X)\le \pur(\psi_E)\label{bound1}
\ea
while we do not know, in principle, whether $\pur(\rho_E)\le \pur(\rho_F)$; 
\begin{proof}
Consider the state $\psi_t$ and its partial state on $E$ i.e., $\psi_E$. We are interested in knowing $\pur(\psi_E)$. In general, the purity of $\psi_E$ is written as
\be
\pur(\psi_E)=\frac{1}{d_E}\sum_{P\in \mathbb{P}_E}\tr^2(P_E\psi_{E})\ge \frac{1}{d_E}\sum_{P_E\,:\,\tr(P_E\psi_E)=\pm 1}
\ee
i.e., is of course lower bounded by the number of Pauli operators on $E$ that have expectation value $1$ in absolute value. Now, let $G\subset \mathbb{P}$ the subset of the Pauli group such that
\be
G=\{P\in\mathbb{P}\,|\, P\ket{\psi_t}=\pm\ket{\psi_t}\}
\ee
taking the partial trace of the set $G$, one create the following set
\be
G_{E}:=\{P_E\in\mathbb{P}_E\,|\, P_E=\tr_{F}(P),\,P\in G\}
\ee
and trivially one has $\frac{1}{d_E}\sum_{P_E\,:\,\tr(P_E\psi_E)=\pm 1}= \frac{|G_E|}{d_E}$. Now, let $W$ be the diagonalizer and let it act on the state $\ket{\psi}$. We know that the diagonalizer takes $G$ to a $WGW^{\dag}\equiv G^{\prime}$. Define $G_{\bar{Y}}\equiv \tr_Y (G^\prime):=\{P_{\bs}\in \mathbb{P}_{\bs}\,|\, P_{\bs}=\tr_{Y}(P), P\in G^{\prime}\}$. Note that, by construction, $G_{\bar{Y}}$ is the same stabilizer group of $\tr_Y(\Phi)\equiv\Phi_{\bs}$. This is because $W\psi W^{\dag}=c_{t}^{Y}\Phi c_{t}^{Y\dag}$ and $\tr_Y(c_{t}^{Y}\Phi c_{t}^{Y\dag})=\tr_Y(\Phi)$. By applying the diagonalizer back $W^{\dag}$, one has that $W^{\dag}G_{\bs}^{\prime}W\in G$, where $G_{\bs}^{\prime}\equiv \{P\in\mathbb{P}\,|\, P=P_{\bs}\otimes \bbbone_Y\,, P_{\bs}\in G_{\bs}\}$; note that $W^{\dag}G_{\bs}^{\prime}W$ is now the stabilizer group of $\rho$; first of all $W^{\dag}G_{\bs}^{\prime}W\subset G$ and, in general, $|W^{\dag}G_{\bs}^{\prime}W|\le |G|$; consequently, defining $\tilde{G}_E\equiv\{P_E\in\mathbb{P}_E\,|\, P_E=\tr_F(P), P\in W^{\dag}G_{\bs}^{\prime}W\}$, we have $|\tilde{G}_{E}|\le |G_E|$. Noting that $\pur(\rho_E)=|\tilde{G}_{E}|/d_E$,  we have the following inequality:
\be
\pur(\rho_E)\le \frac{|G_E|}{d_E}=\frac{1}{d_E}\sum_{P_E\,:\,\tr(P_E\psi_E)=\pm 1}\le \pur(\psi_E)
\ee
The proof is concluded. The bound $\pur(\psi_E)\ge \pur(\rho_F)$ is proven in the same way by noting that $\pur(\psi_E)=\pur(\psi_F)\ge \pur(\rho_F)$.
\end{proof}
\end{proposition}

\begin{proposition}\label{propupperbound}
The purity of the state $\psi_E$ is upper bounded by 
\be
\pur(\psi_E)\le d_{Y}^{2}\pur(\rho_X)\label{upperbound}
\ee
see App~\ref{app:exactupperbound} for the proof.
\end{proposition}
%{\bf Proposition 4.} 
%Let $c_t$ be a $t$-doped Clifford circuit acting on a $t$-qubit subsystem. Then the average purity over the realizations of $c_t$ gives
%\be
%\mathbb{E}_{c_t}[\pur(\psi_E)]\le 2\pur(\rho_X)\label{averagepurityct}
%\ee
%for $\rho_X=\operatorname{argmax}\{\pur(\rho_E),\pur(\rho_F)\}$.

Let us now explain the application of the protocol. Without loss of generality, posit $\pur(\rho_E)>\pur(\rho_F)$ and thus $X\equiv E$. After evaluating $\pur(\rho_E)$, Proposition~\ref{Prop: exactlowerbound} immediately tells us what is a sufficient number of measurements $N_{\text{shot}}$ to resolve the purity of $\psi_E$ by the swap test. In case (i) this is a polynomial number, and thus purity can be efficiently estimated with a polynomial algorithm. In case (ii), recalling that $n_Y\!\!=\!\! t$ and thus $d_{Y}=O(\poly(n))$,  Proposition~\ref{propupperbound} implies that one can estimate the purity as~\footnote{Set $\pur(\rho_E)=\Theta(2^{-\alpha n})$ and $t=O(\log_2 n)$. Asymptotically (in $n$) there exist two constants $A,B$ such that $Be^{-\alpha n}\le \pur(\rho_E)\le Ae^{-\alpha n}$. From Eq.~\eqref{upperbound}, we can thus write $Be^{-\alpha n}\le \pur(\psi_E)\le  A^{\prime}\poly(n)2^{-\alpha n}$. Therefore noticing that $\poly(n)=2^{\log_2\poly n}=2^{O(\log_2 n)}$, we can write $\pur(\psi_E)=2^{-\alpha n+f(n)}$, where $f(n)=O(\log_2 n)$ and $f(n)=\Omega(1)$.}
\be
\pur(\psi_E)=2^{-\alpha n +O(\log_2 n)}\label{Eq.efficientpurityestimation}
\ee
i.e., we estimate the bipartite entanglement in $E|F$ up to a second-order logarithmic correction. This is the second main result of our paper: we can estimate an exponentially small purity by a polynomial number of measurements, thereby achieving an exponential improvement over the known state-of-the-art algorithms.

%by Proposition 4 and the Markov inequality we easily obtain (see~\cite{SeeSupplementalMaterial}) that 
%\be\label{eq:tightbound}
%\pur(\psi_E)=\Theta(2^{-\beta n})
%\ee
%where $\alpha-\delta\le \beta\le \alpha$, 
%and $\delta \ll\alpha$ (although $\delta=\Theta(1)$),
%with probability $1-O(e^{-\delta n})$, for all $\delta>0$. 

{\em Conclusions.---} In this paper, we showed that the stabilizer entropy can be moved around subsystems. Effectively, this results in reducing the complexity of a selected subsystem. The tension between the spreading of non-stabilizerness and its localization is akin to an insulator-superfluid transition. In the localized phase, one can exploit this reduction of complexity in relevant quantum information protocols: we show a way of estimating an exponentially small purity by polynomial resources, thereby improving dramatically on known methods. 

In perspective, there are a number of questions raised by this paper that we find of interest. First of all, the scope of the purity estimation algorithm presented here can be extended to an efficient SE estimation. Similarly, the cleansing algorithm can potentially be utilized as a starting point for a whole family of quantum algorithms aimed at exploiting the easiness of handling stabilizer states even in non-stabilizer settings. 

Then, more generally, how does the complexity cleansing algorithm generalize to time evolution generated by a Hamiltonian? What is the connection between SE cleansing and quantum error-correcting codes? Finally, being an entropy, can SE be evaluated geometrically in the general context of AdS/CFT\cite{white2021ConformalFieldTheories}?

{\em Acknowledgements.---} The authors thank the anonymous Referee for prompting us to significantly improve our algorithm. The authors acknowledge support from NSF award number 2014000. A.H. acknowledges financial support from PNRR MUR project $PE0000023$-NQSTI and PNRR MUR project CN $00000013$ -ICSC. L.L. and S.F.E.O. contributed equally to this paper.
%\let\oldaddcontentsline\addcontentsline% Store \addcontentsline
%\renewcommand{\addcontentsline}[3]{}% Make \addcontentsline a no-op
%\medskip
%\bibliographystyle{apsrev4-2}
%\bibliography{biblio}
\appendix
\setcounter{secnumdepth}{2}
\setcounter{equation}{0}
\setcounter{figure}{0}
\renewcommand{\thetable}{\Alph{section}\arabic{table}}
\renewcommand{\theequation}{\Alph{section}\arabic{equation}}
\renewcommand{\thefigure}{\Alph{section}\arabic{figure}}
\clearpage
\onecolumngrid
\begin{center}
\textbf{\large Appendix}
\end{center}
\setcounter{equation}{0}
\setcounter{figure}{0}
\setcounter{table}{0}

\section{Big-O notation, a brief summary}\label{App: bigO}
In this section, we briefly review: $O(\cdot)$, $\Omega(\cdot)$ and $\Theta(\cdot)$ notations. Consider, for simplicity, positive functions $f(n), g(n)$ of natural numbers $n\in\mathbb{N}$. We can think $n$ as the number of qubits in a multi-qubit quantum system. Then
\be
f(n)=O(g(n))
\ee
if and only if there exists a constant $a\ge 0$ and a certain value $n_{>}\in\mathbb{N}$ such that
\be
\forall n\ge n_{>},\quad f(n)\le a\, g(n)\,.
\ee
Conversely:
\be
f(n)=\Omega(g(n))
\ee
if and only if there exists a constant $b\ge 0$ and a value $n_{<}\in\mathbb{N}$ such that
\be
\forall n\ge n_{<},\quad f(n)\ge b\, g(n)\,.
\ee
Lastly:
\be
f(n)=\Theta(g(n))
\ee
if and only if there exist two constant $c_1,c_2\ge 0$ and a value $n_{=}\in\mathbb{N}$ such that
\be
\forall n\ge n_{=},\quad c_1\,g(n)\le f(n)\le c_2\, g(n)
\ee
Let us give some clarifying examples. Consider $f(n)=10n^3$. We say that $f(n)=O(2^{n})$ because there exists $n_0=15$ after which $f(n)\le 2^n$. We can also write $f(n)=\Omega(n^2)$ because for any $n\ge 1$ $f(n)\ge n^2$. Lastly, we write $f(n)=\Theta(n^3)$ because there exist two constants $c_1=9$, $c_2=11$ such that for any $n\ge 1$, one has $9n^3\le f(n)\le 11n^3$. After the above trivial warm-up example, let us make another last example. Consider $f(n)$ as a sum of two exponentials $f(n)=e^{-2n}+e^{-n}$. We say that $f(n)=\Theta(e^{-n})$ because there exist two constant $c_1=1$ and $c_2=1+e^{-4}$ such that for every $n\ge 2$ one has $e^{-n}\ge f(n)\le (1+e^{-4})e^{-n}$.

\section{Proof of Proposition~\ref{prop1}}\label{App:proofprop1}
In this section, we compute the average value of $w(\psi_{X}^{C}):=\frac{d_X\,\operatorname{SP}(\psi_{X})}{\pur(\psi_{X}^{C})}$, with $X=\{E,F\}$, over the Clifford group. 
\be
\mathbb{E}_C[w(\psi_{E}^{C})]=\mathbb{E}_C\frac{d_E\,\operatorname{SP}(\psi_{E})}{\pur(\psi_{E}^{C})}\,,
\quad
\mathbb{E}_C[w(\psi_{F}^{C})]=\mathbb{E}_C\frac{d_F\,\operatorname{SP}(\psi_{F})}{\pur(\psi_{F}^{C})}\label{app2}
\ee
We would like to calculate the average of the ratio in \eqref{app2} as the ratio of the individual averages. Starting from Lemma~\ref{lemma1} one can then proceed with the evaluation of the averages.
Let us recall a result from~\cite{leone2021QuantumChaosQuantum,oliviero2021TransitionsEntanglementComplexity}: let $\psi$ be a pure quantum state, then its Clifford orbit $\mathbb{E}_{C}[\psi^{C\otimes 4}]$ where $\psi^C\equiv C\psi C^{\dag}$ reads
\be\label{avgclif}
\mathbb{E}_{C}[\psi^{C\otimes 4}]=\alpha Q\Pi_{\sym}+\beta \Pi_{\sym}
\ee
where 
\begin{equation}\label{psym}
   \Pi_{\sym}=\frac{1}{4!}\sum_{\pi\in S_4}T_{\pi} 
\end{equation} and
\ba
\alpha&=\frac{\operatorname{SP}(\psi)}{(d+1)(d+2)/6}-\frac{1-\operatorname{SP}(\psi)}{ (d-1) (d+1) (d+2) (d+4)/24}\\
\beta&=\frac{1-\operatorname{SP}(\psi)}{ (d-1) (d+1) (d+2) (d+4)/24}
\ea
\begin{remark}
Notice that, despite the similar notation, the permutation operators introduced in \eqref{psym} are very different from the permutation operator used to define $W$. The first are a representation over $\mathcal{H}^{\otimes 4}$ of the permutation group of four objects $S_4$, regardless of the size of the system, which act by switching the basis element of the single Hilbert space $\mathcal{H}$ among the other three copies, according to $\pi \in S_4$, whereas the operators $T_{\pi_Y}$ are a representation of the permutation group of n elements over $\mathcal{H}$, that act as simply shuffling the qubits of the system according to the permutation of $S_n$ of choice. 
\end{remark}
Using this knowledge let us evaluate the numerator of Eq.~\eqref{ratioavgE}. Since $\operatorname{SP}(\psi)$ is a linear operator of $\psi^\otimes 4$, we are able to slide the Clifford expectation value into it. One then gets
\ba
 \tr(Q_{E}\mathbb{E}_C[\psi^{C\otimes4}])&=\tr(Q_E\otimes \bbbone_F(\alpha Q\Pi_{\sym}+\beta \Pi_{\sym}))\\&=\alpha\tr(Q\Pi_{\sym})+\beta\sum_{\pi}\tr(Q_ET_{\pi}^{(E)})\tr(T_{\pi}^{(F)})\\
&=\alpha(d+1)(d+2)/6+\beta\sum_{\pi}\tr(Q_ET_{\pi}^{(E)})\tr(T_{\pi}^{(F)})\label{numavgE}
\ea
where $\tr(Q^{(E)}T_{\pi}^{(E)})$ are displayed in~\cite{leone2021QuantumChaosQuantum}. Conversely, the denominator in Eq.~\eqref{ratioavgE} reads~\cite{page1993AverageEntropySubsystem}
\be\label{denavgE}
\mathbb{E}_{C}[\pur(\psi^{C}_{E})]=\frac{(d_{E}+d_{F})}{d_Ed_F+1}
\ee
Taking the ratio between Eqs.~\eqref{numavgE} and~\eqref{denavgE}, we get
\be
\mathbb{E}_C[M_{\operatorname{lin}}(\psi_{E}^C)]=1-\mathbb{E}_C[w(\psi_{E}^C)]=\frac{M_{\operatorname{lin}}(\psi)(d_E^2-1)d}{(d-1)(d+d_E^2)}
\ee
in the limit of $n_E\gg n_F$:
\be
\mathbb{E}_C[M_{\operatorname{lin}}(\psi_{E}^C)]=M_{\operatorname{lin}}(\psi)+O\left(\frac{d_F}{d_E}\right)\,.
\ee
In a similar fashion, one could evaluate the other ratio in Eq.~\eqref{ratioavgF}. Recall that, thanks to the Schmidt decomposition, one has $\mathbb{E}_{C}(\pur(\psi_F^C))=\mathbb{E}_{C}(\pur(\psi_E^C))=\frac{d_E+d_F}{d_Ed_F+1}$. 
The numerator of Eq.~\eqref{ratioavgF} reads:
\newline 
\ba
\tr(Q_{F}\mathbb{E}_C[\psi^{C\otimes4}])&=\alpha\tr(Q\Pi_{\sym})+\beta\sum_{\pi}\tr(Q_FT_{\pi}^{(F)})\tr(T_{\pi}^{(E)})\\&=\alpha(d+1)(d+2)/6+\beta\sum_{\pi}\tr(Q_FT_{\pi}^{(F)})\tr(T_{\pi}^{(E)})\label{numavgF}
\ea
by taking the ratio between Eqs.~\eqref{numavgF} and~\eqref{denavgE}, we get
\be
\mathbb{E}_C[M_{\operatorname{lin}}(\psi_{F}^C)]=1-\mathbb{E}_C[w(\psi_{F}^C)]=\frac{M_{\operatorname{lin}}(\psi)(d^2-d_E^2)}{(d-1)(d+d_E^2)}
\ee
in the limit of $n_E\gg n_F$, one gets
\be
\mathbb{E}_C[M_{\operatorname{lin}}(\psi_{F}^C)]=O\left(\frac{d_F}{d_E}\right)
\ee
which proves Proposition \ref{prop1}.\qed

\section{Proof of Eq.~\eqref{eq:m2g}}\label{App:average}

\subsection{Average over the cleansing map}\label{App: averagecleansing}
In this section, we show that averaging over the maps $\mathcal{E}\circ \mathcal{C}_{\mathfrak{t}}(\omega)$ denoted as $\mathbb{E}$ is equivalent to averaging over the doped Clifford circuits $C_t$. First note that $C_t=D^{\dag}c_t DV$; then thanks to the left or right invariance of the Haar measure one has the average over $t$-doped Clifford circuits $\mathbb{E}_{C_t}$ introduced in~\cite{leone2021QuantumChaosQuantum} equals $\mathbb{E}_{C_t}=\mathbb{E}_{V}\mathbb{E}_{D}\mathbb{E}_{c_t}$, where $c_t$ is a $t$-doped Clifford on $t$ qubits and $D,V$ belong to the  Clifford group. Finally, note that $\mathcal{E}\circ \mathcal{C}_{\mathfrak{t}}(\omega)\equiv \tr_Y(WC_t(\omega) C_{t}^{\dag}W^{\dag})=\tr_Y \left(c_t^Y T_{\pi_Y}DV(\omega) (c_t^Y T_{\pi_Y}DV)^{\dag}\right)$. Consequently, the average over $\mathcal{E}\circ \mathcal{C}_{\mathfrak{t}}(\omega)$, hereby denoted as $\mathbb{E}$, of a function $f(\mathcal{E}\circ \mathcal{C}_{\mathfrak{t}}(\omega))$ obey $\mathbb{E}f(\mathcal{E}\circ \mathcal{C}_{\mathfrak{t}}(\omega))\equiv \mathbb{E}_{c_t^Y}\mathbb{E}_{D}\mathbb{E}_{V}\mathbb{E}_{\omega}f(\mathcal{E}\circ \mathcal{C}_{\mathfrak{t}}(\omega))\equiv \mathbb{E}_{c_t}\mathbb{E}_{D}\mathbb{E}_{V} f(\mathcal{E}\circ \mathcal{C}_{\mathfrak{t}}(\omega))\equiv \mathbb{E}_{C_t}f(\mathcal{E}\circ \mathcal{C}_{\mathfrak{t}}(\omega))$, where we used the invariance of the Haar measure over the Clifford group. Note that
\be
\mathbb{E}\,  \left[M_{E}(\mathcal{E}\circ \mathcal{C}_{\mathfrak{t}}(\omega))\right]\ge -\log_2 \mathbb{E}\left[\frac{\operatorname{SP}(\tr_{F\setminus Y}\mathcal{E}\circ \mathcal{C}_{\mathfrak{t}}(\omega))}{\pur(\tr_{F\setminus Y}\mathcal{E}\circ \mathcal{C}_{\mathfrak{t}}(\omega))}\right]
\ee
where $\mathbb{E}$ is the average over $\mathcal{E}\circ \mathcal{C}_{\mathfrak{t}}(\omega)$ discussed in Sec.~\ref{App: averagecleansing}. 
The object we intend to calculate is 
\ba
     \mathbb{E}_{c_t}\mathbb{E}_D \mathbb{E}_\omega  SP(\tr_{F\setminus Y}\mathcal{E}\circ \mathcal{C}_{\mathfrak{t}}(\omega))&=\mathbb{E}_{c_t}\mathbb{E}_D \mathbb{E}_V\tr(Q_E\otimes\bbbone_F^{\otimes 4})(c_t^{Y^{\otimes 4}}\otimes\bbbone_{\bar{Y}}^{\otimes 4})\\&\times(T_{\pi_Y}DV)^{\otimes 4}\ket{\mathbf{0}}\bra{\mathbf{0}}^{\otimes 4}(T_{\pi_Y}DV)^{\dagger\otimes 4}(c_t^{Y^{\otimes 4}}\otimes\bbbone_{\bar{Y}}^{\otimes 4})^\dagger  
\ea
Where $\bar{Y}=(E\cup F)\cap Y$. Due to the left and right unitary invariance of the Haar measure over the Clifford group, the average over $D$ gets absorbed into the one over $V$, and the permutation operator gets absorbed as well. One then gets:
\ba
        \mathbb{E}_{c_t}\mathbb{E}_D \mathbb{E}_V  SP(\tr_{F\setminus Y}\mathcal{E}\circ \mathcal{C}_{\mathfrak{t}}(\omega))&=\mathbb{E}_{c_t}\mathbb{E}_V  SP(\tr_{F\setminus Y}\mathcal{E}\circ \mathcal{C}_{\mathfrak{t}}(\omega))=\tr(Q_E\otimes\bbbone_F^{\otimes 4})\mathbb{E}_{c_t}(c_t^{Y^{\otimes 4}}\otimes\bbbone_{\bar{Y}}^{\otimes 4})\\&\times\mathbb{E}_V V^{\otimes 4}\ket{\mathbf{0}}\bra{\mathbf{0}}^{\otimes 4}V^{\dagger\otimes 4}(c_t^{Y^{\otimes 4}}\otimes\bbbone_{\bar{Y}}^{\otimes 4})^\dagger  
\ea
The average over V gives the Clifford $n$-qubit state orbit, as shown in \eqref{avgclif}. By substituting the expression we get
\be
    \mathbb{E}_{c_t}\mathbb{E}_V  SP(\tr_{F\setminus Y}\mathcal{E}\circ \mathcal{C}_{\mathfrak{t}}(\omega))=\frac{1}{4!} \sum_\pi\tr(Q_E\otimes\bbbone_F^{\otimes 4})\mathbb{E}_{c_t}(c_t^{Y^{\otimes 4}}\otimes\bbbone_{\bar{Y}}^{\otimes 4})(\alpha Q T_\pi+\beta T_\pi)(c_t^{Y^{\otimes 4}}\otimes\bbbone_{\bar{Y}}^{\otimes 4})^\dagger 
\ee
Here we will exploit the fact that the permutation operators $T_\pi$ and $Q$ in $\mathcal{B}(\mathcal{H}^{\otimes 4})$  can always be factorized as $T_\pi^{(Y)}\otimes T_\pi^{(\bar{Y})}$ and $Q_Y\otimes Q_{\bar{Y}}$ with $T_\pi^{(Y)},Q_Y \in \mathcal{B}(\mathcal{H}_Y^{\otimes 4})$ and $T_\pi^{(\bar{Y})},Q_{\bar{Y}} \in \mathcal{B}(\mathcal{H}_{\bar{Y}}^{\otimes 4})$, being $\mathcal{H}=\mathcal{H}_Y\otimes\mathcal{H}_{\bar{Y}}$. In this way we can write
\ba\label{wiz}
        \mathbb{E}_{c_t}\mathbb{E}_V  SP(\tr_{F\setminus Y}\mathcal{E}\circ\mathcal{C}_t(\omega))&=\frac{1}{24} \sum_\pi\tr(Q_E\otimes\bbbone_F^{\otimes 4})\mathbb{E}_{c_t}(c_t^{Y^{\otimes 4}}\otimes\bbbone_{\bar{Y}}^{\otimes 4})(\alpha Q_Y T^{(Y)}_\pi\otimes Q_{\bar{Y}}T^{\bar{(Y)}}_\pi
        \\
        &+\beta T^{(Y)}_\pi\otimes T^{\bar{(Y)}}_\pi)(c_t^{Y^{\otimes 4}}\otimes\bbbone_{\bar{Y}}^{\otimes 4})^\dagger\\
        &=\frac{1}{24} \sum_\pi\tr(Q_E\otimes\bbbone_F^{\otimes 4})\Big(\alpha\mathbb{E}_{c_t}( c_t^{Y^{\otimes 4}}Q_Y T_\pi^{(Y)}c_t^{Y\dagger^{\otimes 4}})\otimes Q_{\bar{Y}}T^{\bar{(Y)}}_\pi \\
        &+\beta \mathbb{E}_{c_t}( c_t^{Y^{\otimes 4}} T_\pi^{(Y)} c_t^{Y\dagger^{\otimes 4}})\otimes T^{\bar{(Y)}}_\pi\Big)
        \\
        &\equiv R_1+ R_2
\ea
As before, due to the left and right invariance of the Haar measure over the Clifford group, the average over $c_t$ of $c_t^Y$ is the same of the average of $c_t$ over $c_t$, since the permutation operator which dresses $c_t$ is a Clifford operator.
%\begin{equation}
     %\frac{1}{24}\sum_\pi\tr{(Q_E\otimes\bbbone_F)\left(\alpha\mathbb{E}_{C}\tilde{U_t}^{\otimes 4}Q_t T_\pi^{(t)}\tilde{U_t}^{\otimes 4}_{\tilde{U_t}} \otimes Q_{Y}T_\pi^{(Y)}+\\ \beta_n\mathbb{E}_{C}\tilde{U_t}^{\otimes 4}T_\pi^{(t)}\tilde{U_t}^{\otimes 4}_{\tilde{U_t}} \otimes T_\pi^{(Y)}\right)}
%\end{equation}
   The way to calculate the averages over $c_t$ is shown in \cite{leone2021QuantumChaosQuantum} and the result reads
\ba
           \mathbb{E}_{c_t}c_t^{Y^{\otimes 4}}Q_Y T_\pi^{(Y)}c_t^{Y\dagger^{\otimes 4}}&=\sum_\sigma\eta_\sigma(Q_Y T_\pi^{(Y)})Q_Y T_\sigma^{(Y)}+\mu_\sigma(Q_Y T_\pi^{(Y)})T_\sigma^{(Y)}\\
             \mathbb{E}_{c_t}c_t^{Y^{\otimes 4}}T_\pi^{(Y)}c_t^{Y\dagger^{\otimes 4}}&=\sum_\sigma\eta_\sigma( T_\pi^{(Y)})Q_Y T_\sigma^{(Y)}+\mu_\sigma( T_\pi^{(Y)})T_\sigma^{(Y)}
\ea
    with
\ba
           \eta_\sigma(\mathcal{O})&=\sum_{\pi}\Xi_{\sigma\pi}^{t} c_\pi(\mathcal{O})\\
           \mu_\sigma(\mathcal{O})&=b(\mathcal{O})+\sum_{\pi}\Gamma_{\sigma\pi}^{(t)}c_\pi(\mathcal{O})
\ea
   And $c_\pi,b_\pi,\Xi_{\pi\sigma}$ and $\Gamma_{\pi\sigma}^{(t)}$ as defined in \cite{leone2021QuantumChaosQuantum}.
   Since both the averages have non-zero components both on $Q_Y T_\sigma^{(Y)}$ and $T_\sigma^{(Y)}$, the terms $R_1, R_2$ in \eqref{wiz} will have the same structure, with the only difference being the value of the coefficients $\eta_\sigma$ and $\mu_\sigma$, so we will carry on the calculation for just the first one:
   \begin{equation}\label{calcstruct}
       \begin{split}
          R_1= \frac{\alpha}{24}\sum_{\pi\sigma}\tr\left(\eta_\sigma(Q_Y T_\pi^{(Y)})Q_Y T_\sigma^{(Y)}+\mu_\sigma(Q_Y T_\pi^{(Y)})T_\sigma^{(Y)}\right)\otimes Q_{\bar{Y}}T_\pi^{\bar{(Y)}}(Q_E\otimes\bbbone_F)
       \end{split}
   \end{equation}
   Since $\mathfrak{t}/\mathfrak{f}>1$ we can write the partition $E$ as $E=E\cup F\cap Y\cup Y\cap F=\bar{Y}\cup Y\cap F= \bar{Y}\cup Y'$
with $Y'=Y\cap F$. In the same fashion, we can write $Y=F\cup Y'$ and factorize the $Q$ and $T$ operators accordingly as
\begin{equation}
\begin{split}
   Q_E&=Q_{\bar{Y}}\otimes Q_{Y'}\\
   T_\pi^{(E)}&= T_\pi^{\bar{(Y)}}\otimes  T_\pi^{(Y')}\\
   Q_Y&=Q_{F}\otimes Q_{Y'}\\
    T_\pi^{(Y)}&= T_\pi^{(F)}\otimes  T_\pi^{(Y')}
\end{split}
\end{equation}
Substituting into \eqref{calcstruct} one gets
  \ba
          R_1 &=\frac{\alpha}{24}\sum_{\pi\sigma}\tr{\left(\eta_\sigma(Q_Y T_\pi^{(Y)})Q_Y T_\sigma^{(Y)}+\mu_\sigma(Q_Y T_\pi^{(Y)})T_\sigma^{(Y)}\right)\otimes Q_{\bar{Y}}T_\pi^{\bar{(Y)}}(\bbbone_F\otimes Q_{Y'}\otimes Q_{Y})}\\ 
           &=\frac{\alpha}{24}\sum_{\pi\sigma}\tr{\left(\eta_\sigma(Q_Y T_\pi^{(Y)})Q_Y T_\sigma^{(Y)}+\mu_\sigma(Q_Y T_\pi^{(Y)})T_\sigma^{(Y)}\right)(\bbbone_F\otimes Q_{Y'})}\tr(Q_{\bar{Y}}T_\pi^{\bar{(Y)}})\\ 
           &=\frac{\alpha}{24}\sum_{\pi\sigma}\tr\left(\eta_\sigma(Q_Y T_\pi^{(Y)})Q_{F} T_\sigma^{(F)}\otimes Q_{Y'} T_\sigma^{(Y')}+\mu_\sigma(Q_{Y} T_\pi^{(Y)})T_\sigma^{(F)}\otimes T_\sigma^{(Y')}\right)(\bbbone_F\otimes Q_{Y'})\tr(Q_{\bar{Y}}T_\pi^{\bar{(Y)}})\\
           &=\frac{\alpha}{24}\sum_{\pi\sigma}\tr(\eta_\sigma(Q_Y T_\pi^{(Y)})Q_Y T_\sigma^{(Y)}+\mu_\sigma(Q_Y Y_\pi^{(Y)})Q_{Y'}T_\sigma^{(Y')}\otimes T_{\sigma}^{F})\tr(Q_{\bar{Y}}T_\pi^{\bar{(Y)}})\\  
           &=\frac{\alpha}{24}\sum_{\pi\sigma}\eta_\sigma(Q_Y T_\pi^{(Y)})\tr(Q_Y T_\sigma^{(Y)})\tr(Q_{\bar{Y}}T_\pi^{\bar{(Y)}})+
           \mu_\sigma(Q_Y T_\pi^{(Y)})\tr(Q_{Y'}T_\sigma^{(Y')})\tr(T_\sigma^{(F)})\tr(Q_{\bar{Y}}T_\pi^{\bar{(Y)}})
  \ea
     By plugging the values of the coefficients $\eta$ and $\mu$ computing the traces one gets the expression of the function $g(n,\mathfrak{t},\mathfrak{f})$ (fully displayed in the following section).\qed
   \subsection{Explicit formula for the partial SE in E}
   In this section, we show the full-expression for $g(n,\mathfrak{t},\mathfrak{f})$,
   \begin{equation}
       \begin{split}
           g(n,\mathfrak{t},\mathfrak{f})&=\left({3 (d+2) (d+4) (d+2^{2 n(1-\mathfrak{f})}) (2^{2 n\mathfrak{t}}-9)}\right)^{-1}\\
           &\times\bigg(2^{-6 n\mathfrak{t}-1} \Big(2^{2 n(1-\mathfrak{f})+4 n\mathfrak{t}+1} \big(2^{n\mathfrak{t}} (3\ 2^{2 n\mathfrak{t}} f_-^{n\mathfrak{t}}+2^{3 n\mathfrak{t}} f_-^{n\mathfrak{t}}-5\ 2^{n\mathfrak{t}+1} f_-^{n\mathfrak{t}}-24 f_-^{n\mathfrak{t}}+\\
           &+(2^{n\mathfrak{t}}-4) (2^{n\mathfrak{t}}-2) (2^{n\mathfrak{t}}+3) f_+^{n\mathfrak{t}}-9\ 2^{n\mathfrak{t}+4}+9\ 2^{3n\mathfrak{t} +1})+\\
           &-2 (-13\ 2^{2 {n\mathfrak{t}}}+2^{4 {n\mathfrak{t}}}+36) g^{n\mathfrak{t}}\big)+2^{2 n+2 {n\mathfrak{t}}} \big(2^{4 {n\mathfrak{t}}} (2^{2 n(1-\mathfrak{f})} (f_-^{n\mathfrak{t}}+f_+^{n\mathfrak{t}}+18)+\\&-2 (f_-^{n\mathfrak{t}}+f_+^{n\mathfrak{t}}-2 g^{n\mathfrak{t}}))-2^{2 {n\mathfrak{t}}+1} (2^{2 n(1-\mathfrak{f})} (5 f_-^{n\mathfrak{t}}+5 f_+^{n\mathfrak{t}}+24 g^{n\mathfrak{t}}+72)+\\
           &-10 f_-^{n\mathfrak{t}}-10 f_+^{n\mathfrak{t}}+26 g^{n\mathfrak{t}}+228)+3 (2^{2 n(1-\mathfrak{f})}-2) 2^{3 {n\mathfrak{t}}} (f_-^{n\mathfrak{t}}-f_+^{n\mathfrak{t}})+\\
           &-3 (2^{2 n(1-\mathfrak{f})}-2) 2^{{n\mathfrak{t}}+3} (f_-^{n\mathfrak{t}}-f_+^{n\mathfrak{t}})+144 (2^{2 n(1-\mathfrak{f})}+1) g^{n\mathfrak{t}}+9\ 2^{6 {n\mathfrak{t}}+2}\big)+\\
           &+2^{4 n} \big(2^{n\mathfrak{t}} (24 (f_-^{n\mathfrak{t}}-f_+^{n\mathfrak{t}})+2^{n\mathfrak{t}} (3\ 2^{n\mathfrak{t}} (f_+^{n\mathfrak{t}}-f_-^{n\mathfrak{t}})-2^{2 {n\mathfrak{t}}} (f_-^{n\mathfrak{t}}+f_+^{n\mathfrak{t}}+72)+\\
           &+2 (5 f_-^{n\mathfrak{t}}+5 f_+^{n\mathfrak{t}}+72)+3\ 2^{4 {n\mathfrak{t}}+1}))+48 (2^{2 {n\mathfrak{t}}}-3) g^{n\mathfrak{t}}\big)+\\
           &+3 \big(24 (f_-^{n\mathfrak{t}}+f_+^{n\mathfrak{t}})+2^{n\mathfrak{t}} (-3\ 2^{n\mathfrak{t}} f_-^{n\mathfrak{t}}-2^{2 {n\mathfrak{t}}} f_-^{n\mathfrak{t}}+10 f_-^{n\mathfrak{t}}+\\
           &+(2^{n\mathfrak{t}}-5) (2^{n\mathfrak{t}}+2) f_+^{n\mathfrak{t}}-15\ 2^{{n\mathfrak{t}}+2}+2^{5 {n\mathfrak{t}}+1})\big) 2^{3 n+2 {n\mathfrak{t}}}+3\ 2^{n+4 {n\mathfrak{t}}} t\\
           &\times\big(-5 (f_-^{n\mathfrak{t}}-f_+^{n\mathfrak{t}}) 2^{2 n(1-\mathfrak{f})+{n\mathfrak{t}}+1}+(f_-^{n\mathfrak{t}}-f_+^{n\mathfrak{t}}) 2^{2 n(1-\mathfrak{f})+3 {n\mathfrak{t}}}+\\&-3\ 2^{2 n(1-\mathfrak{f})+3} (f_-^{n\mathfrak{t}}+f_+^{n\mathfrak{t}})+3\ 2^{2 {n\mathfrak{t}}} (2^{2 n(1-\mathfrak{f})} (f_-^{n\mathfrak{t}}+f_+^{n\mathfrak{t}}-20)-48)+\\
           &+(3\ 2^{2 n(1-\mathfrak{f})}+8) 2^{4 {n\mathfrak{t}}+1}\big)\Big)\bigg)\,,
       \end{split}
   \end{equation}
   with
   \begin{equation}
       f_{-}=\frac{3\times 4^{n\mathfrak{t}}-3\times 2^{n\mathfrak{t}}-4}{4^{n\mathfrak{t}}-1},
f_{+}=\frac{3\times 4^{n\mathfrak{t}}+3\times 2^{n\mathfrak{t}}-4}{4^{n\mathfrak{t}}-1},
g=\frac{3\times 4^{n\mathfrak{t}}-4}{4^{n\mathfrak{t}}-1}\,.
   \end{equation}
\section{Efficient Purity Estimation}

Let us first of all establish some useful notation. Recalling the definition for the stabilizer state $\ket{\Phi}: =  WV\ket{\mathbf{0}}$ and $\Phi\equiv\st{\Phi}$,  we obtain  the identity $\tr_{Y} (W \psi_t W^{\dag})= \tr_{Y}(\Phi)\equiv\Phi_{\bs}$ and notice that this is a stabilizer state. Notice that $\mathcal{E}(\psi_t)=\Phi_{\bs}$ and thus the stabilizer state $\rho =W^{\dag} (\mathcal{E}(\psi_t)\otimes d_{Y}^{-1}I_{Y})W=W^{\dag} (\Phi_{\bs}\otimes d_{Y}^{-1}I_{Y})W$.   %In the following, we show that through the purities of $\rho_E,\rho_F$ we can access $\pur(\psi_E)$. %The intuition is that, in the localized phase, the stabilizer $\rho$ contains valuable information about the bipartite entanglement in $\psi_t$.

\subsection{Proof of Proposition 4}\label{app:exactupperbound}
Following the notations of the main paper, let us denote $\ket{\Phi}=WV\ket{\bold{0}}$ and let $\ket{\psi_t}=W^{\dag}c_{t}^{Y}WV\ket{\bold{0}}$ with $W$ being the diagonalizer and $c_{t}^{Y}$ being a $t$-doped Clifford circuit acting on the system $Y$ with $n_Y\le t$. Let us prove the upper bound in Proposition $4$, i.e.,
\be
\pur(\psi_E)\le d_{Y}^{2}\pur(\rho_X)
\ee
for $X=E,F$ and $\rho=W^{\dag}(\Phi_{\bar{Y}}\otimes d_{Y}^{-1}I_Y)W$ and $\Phi_{\bar{Y}}=\tr_Y\st{\Phi}$.

Define $S_E=\{P\in\mathbb{P}\,|\, P=WP_EW^{\dag}\}$. First note that
\be
\pur(\psi_E)=\frac{1}{d_E}\sum_{P_E\in\mathbb{P}_E}\tr^2(\psi_E P_E)=\frac{1}{d_E}\sum_{P\in S_E}\tr^2(c_{t}\Phi c_{t}^{\dag}P)
\ee
Then, thanks to their tensor product structure, Pauli operators $S_E\ni P$ can be decomposed over $Y\cup \bar{Y}$. Therefore, let us define $S_{E}|_{\bar{Y}}=\{d_Y\sum_{P_Y}\tr_Y(P_Y P)\,|\, P\in S_E\}$ the restriction of $S_E$ to $\bar{Y}$. Define the completion of $S_{E}|_{\bar{Y }}$. For $P_{\bar{Y}}\in S_{E}|_{\bar{Y }}$, define the set $T_{P_{\bar{Y}}}=\{d_{\bar{Y}}\tr_{\bar{Y}}(PI_Y\otimes P_{\bar{Y}})\,|\, P\in S_E\}$.
Notice that $S_E$ can be written as
\be
S_E=\bigcup_{P_{\bar{Y}}\in S_{E}|_{\bar{Y}}} \{T_{P_{\bar{Y}}}\otimes P_{\bar{Y}}\}
\ee
We can thus write 
\be
\pur(\psi_E)=\frac{1}{d_E}\sum_{P_{\bar{Y}}\in S_{E}|_{\bar{Y}}}\sum_{P_{Y}\in T_{P_{\bar{Y}}}}\tr^2(c_{t}\Phi c_{t}^{\dag}P_Y\otimes P_{\bar{Y}})= \frac{1}{d_E}\sum_{P_{\bar{Y}}\in S_{E}|_{\bar{Y}}}\sum_{P_{Y}\in T_{P_{\bar{Y}}}}\tr^2(\Phi c_{t}^{\dag}P_Yc_{t}\otimes P_{\bar{Y}})
\ee
Since there is a sum of positive terms, we can upper bound it as
\be
\pur(\psi_E)\le \frac{1}{d_E}\sum_{P_{\bar{Y}}\in S_{E}|_{\bar{Y}}}\sum_{P_{Y}\in \mathbb{P}_Y}\tr^2(\Phi c_{t}^{\dag}P_Yc_{t}\otimes P_{\bar{Y}})=\frac{d_Y}{d_E}\sum_{P_{\bar{Y}}\in S_{E}|_{\bar{Y}}}\tr[\Phi^{\otimes 2}T_{Y}(I_{Y}\otimes P_{\bar{Y}})^{\otimes 2}]
\ee
Now, note that $\tr[\Phi^{\otimes 2}T_{Y}(I_{Y}\otimes P_{\bar{Y}})^{\otimes 2}]=\tr_{\bar{Y}}[\Phi^{\otimes 2}T_{\bar{Y}}(I_{Y}\otimes P_{\bar{Y}})^{\otimes 2}]=\tr_{\bar{Y}}(\Phi_{\bar{Y}}^{\otimes 2} T_{\bar{Y}}P_{\bar{Y}}^{\otimes 2})=\frac{1}{d_{Y}}\tr[\Phi_{\bar{Y}}^{\otimes 2} T_{\bar{Y}}(I_{Y}\otimes P_{\bar{Y}})^{\otimes 2}]$. Where the equality follows from the fact that $T\ket{\Phi^{\otimes 2}}=\ket{\Phi^{\otimes 2}}$. Therefore, we arrived to
\be
\pur(\psi_E)\le \frac{1}{d_E}\sum_{P_{\bar{Y}}\in S_{E}|_{\bar{Y}}}\tr[\Phi_{\bar{Y}}^{\otimes 2} T_{\bar{Y}}(I_{Y}\otimes P_{\bar{Y}})^{\otimes 2}]=\frac{1}{d_E}\sum_{P_{\bar{Y}}\in S_{E}|_{\bar{Y}}}\sum_{P_{Y}\in T_{P_{\bar{Y}}}}\tr[\Phi_{\bar{Y}}^{\otimes 2} T_{\bar{Y}}(P_{Y}\otimes P_{\bar{Y}})^{\otimes 2}]
\ee
the second equality follows from the fact that $\tr[\Phi_{\bar{Y}}^{\otimes 2} T_{\bar{Y}}(P_{Y}\otimes P_{\bar{Y}})^{\otimes 2}]=0$ for every $P_Y\neq I_Y$. Therefore, we have the bound
\be\label{s46}
\pur(\psi_E)\le \frac{1}{d_E}\sum_{P\in S_E}\tr(\Phi_{\bar{Y}}^{\otimes 2}T_{\bar{Y}}P^{\otimes 2})=\tr(W^{\otimes 2}T_E W^{\dag\otimes 2}\Phi_{\bar{Y}}^{\otimes 2}T_{\bar{Y}})=d_{Y}(d_{Y}^{2}-1)\Lambda_1\,,
\ee
with 
\be 
    \Lambda_1:=\frac{1}{d_{Y}(d_{Y}^2-1)}\tr(W^{\dag\otimes 2}T_{E}W^{\otimes 2}\Phi_{\bs}^{\otimes2} T_{\bs})
\ee
The term $\Lambda_1$ is proportional to $\pur(\rho_F)$. First, by evaluating $\pur(\rho_F)$, which reads

\ba
    \pur(\rho_F)&=\frac{1}{d_Y^2}\tr(W^{\dag\otimes2}T_EW^{\otimes 2} \Phi_{\bs}^{\otimes2}T)=\frac{1}{d_Ed_Y^2}\sum_{P_E}\tr(P_E(W)\Phi_{\bs}P_E(W)\Phi_{\bs})\\
    &=\frac{1}{d_Ydd_E}\sum_{P_E,Q_{\bs}}\tr(P_E(W)Q_{\bs}P_{E}(W)\Phi_{\bs})\tr(Q_{\bs}\Phi_{\bs})=\frac{1}{d^2d_E}\sum_{P_E,Q_{\bs}}\tr(P_E(W)Q_{\bs}P_{E}Q_{\bs})\,,
\ea
 we now show the proportionality between $\pur (\rho_F)$ and $\Lambda_1$:

 \ba 
    \Lambda_1&=\frac{1}{d_Y(d_Y^2-1)}\tr(W^{\dag\otimes 2}T_E W^{\otimes 2}\Phi_{\bs}^{\otimes 2}T_{\bs})=\frac{1}{d_Yd_E(d_Y^2-1)}\sum_{P_E}\tr(W^{\dag\otimes 2}P_E^{\otimes 2}W^{\otimes 2}\Phi_{\bs}^{\otimes 2}T_{\bs})\\
    &=\frac{1}{d_Yd_E(d_Y^2-1)}\sum_{P_E}\tr_{\bs}(\tr_Y(P_E(W))\Phi_{\bs}\tr_Y(P_E(W))\Phi_{\bs})\\
    &=\frac{1}{d_Ed(d_Y^2-1)}\sum_{P_E,Q_{\bs}}\tr_{\bs}(\tr_Y(P_E(W))Q_{\bs}\tr_Y(P_E(W))\Phi_{\bs})\tr(Q_{\bs}\Phi_{\bs})\\
    &=\frac{1}{d_{\bs}d_Ed(d_Y^2-1)}\sum_{P_E,Q_{\bs}}\tr_{\bs}(\tr_Y(P_E(W))Q_{\bs}\tr_Y(P_E(W))Q_{\bs})\\
    &=\frac{d_Y}{d_Ed^2(d_Y^2-1)}\sum_{P_E,Q_{\bs}}\tr(P_E(W)Q_{\bs}P_E(W)Q_{\bs})\\
    &=\frac{d_Y}{(d_Y^2-1)}\pur(\rho_F)
 \ea

Therefore, recalling the inequality in Eq. \eqref{s46},
we proved that
\be
    \pur(\psi_E)\le d_{Y}^{2}\pur(\rho_F)
\ee
Similarly, one can obtain the bound for 
\be
    \pur(\psi_F)\le d_{Y}(d_{Y}^{2}-1)\Lambda_2\,,
\ee
with
\be
    \Lambda_2:=\frac{1}{d_Y(d_Y^2-1)}\tr(W^{\dag\otimes 2}T_E W^{\otimes 2}\Phi_{\bs}^{\otimes 2}T_{Y})\,.
\ee 
In a similar fashion to $\Lambda_1$, the term $\Lambda_2$ is proportional to $\pur(\rho_E)$. The latter reads

\ba
    \pur(\rho_E)&=\frac{1}{d_{Y}^{2}}\tr(W^{\dag\otimes2}T_EW^{\otimes 2} \Phi_{\bs}^{\otimes2})=\frac{1}{d_{Y}^{2}}\tr(T W^{\dag\otimes2}T_FW^{\otimes2}\Phi_{\bs}^{\otimes 2})\\
&=\frac{1}{d_{Y}^{2}d_F}\sum_{P_F}\tr(P_F(W)\Phi_{\bs}P_F(W)\Phi_{\bs})=\frac{1}{d_{Y}dd_F}\sum_{P_F,Q_{\bs}}\tr(P_F(W)Q_{\bs}P_{F}(W)\Phi_{\bs})\tr_{\bs}(Q_{\bs}\Phi_{\bs})\\
&=\frac{1}{d^2d_F}\sum_{P_F,Q_{\bs}}\tr(P_F(W)Q_{\bs}P_{F}(W)Q_{\bs})\,,
\ea
 and we can show the proportionality between $\pur(\rho_E)$ and $\Lambda_2$ as follows

 \ba
\Lambda_2&=\frac{1}{d_Y(d_Y^2-1)}\tr(W^{\dag\otimes 2}T_E W^{\otimes 2}\Phi_{\bs}^{\otimes 2}T_{Y})=\frac{1}{d_Y(d_Y^2-1)}\tr(W^{\dag\otimes2}T_FW^{\otimes2}\Phi_{\bs}^{\otimes2}T_{\bs})\\
&=\frac{1}{d_Yd_F(d_Y^2-1)}\sum_{P_F}\tr(W^{\dag\otimes 2}P_FW^{\otimes 2}\Phi_{\bs}^{\otimes 2}T_{\bs})\\
&=\frac{1}{d_Yd_F(d_Y^2-1)}\sum_{P_F}\tr_{\bs}(\tr_Y(P_F(W))\Phi_{\bs}\tr_Y(P_F(W))\Phi_{\bs})\\
&=\frac{1}{d_Fd(d_Y^2-1)}\sum_{P_F,Q_{\bs}}\tr_{\bs}(\tr_Y(P_F(W))Q_{\bs}\tr_Y(P_F(W))\Phi_{\bs})\tr_{\bs}(Q_{\bs}\Phi_{\bs})\\
&=\frac{1}{d_Fd(d_Y^2-1)d_{\bs}}\sum_{P_F,Q_{\bs}}\tr_{\bs}(\tr_Y(P_F(W))Q_{\bs}\tr_Y(P_F(W))Q_{\bs})\\
&=\frac{1}{d_Fd(d_Y^2-1)d_{\bs}}\sum_{P_F,Q_{\bs}}\tr(P_F(W)Q_{\bs}P_F(W)Q_{\bs})\\
&=\frac{d_Y}{d_Fd^2(d_Y^2-1)}\sum_{P_F,Q_{\bs}}\tr(P_F(W)Q_{\bs}P_F(W)Q_{\bs})\\
&=\frac{d_Y}{(d_Y^2-1)}\pur(\rho_E)
\ea

Finally, since $\psi_t$ is pure, one has $\pur(\psi_E)=\pur(\psi_F)$, hence one obtains $\pur(\psi_E)\le d_{Y}^{2}\pur(\rho_E)$, thus concluding the proof. 
\twocolumngrid
%\bibliographystyle{apsrev4-1}
%\bibliography{biblio}

\end{document}